# The 5D to 4D projection model applied as a Lepton to Galaxy Creation model.

Kai-Wai Wong[1], Gisela A.M. Dreschhoff[1], Högne J.N. Jungner

[1]Department of Physics and Astronomy, University of Kansas, Lawrence, Kansas, USA
[2]Radiocarbon Dating Lab, University of Helsinki, Finland

The 5D to 4D projection is presented in a simple geometry giving the Perelman Theorem, resulting in a 3D doughnut structure for the space manifold of the Lorentz space-time. It is shown that in the lowest quantum state, this Lorentz manifold confines and gives the de Broglie leptons from the massless 5D e-trinos. On the scale of the universe, it allows for a model for the creation of galaxies.

## 1. Introduction

In our recent work we have discussed the physics that comes from the 5D homogeneous space-time metric, and the projection actions that would produce the Lorentz 4D space-time metric [1;2]. Many well-known results in physics could then be obtained. This includes the SU(3) representations of hadrons, together with their quantitative masses. However, there are significant differences of the 5D projection theory to those of current conventional theories, and these are part of the discussion in this paper, namely:

(1). The SU(2) and SU(3) symmetries to the elementary particles are the result of projections of the homogeneous 5D space-time manifold, not super-symmetries.
(2). The unit charge 'e' is the coupling constant between massless vector and spinor fields, solutions of the 5D metric operator, not a fundamental parameter, neither is Planck's constant, as postulated in conventional theories.
(3). Riemannian geometry in General Relativity is a result of superposition of the time shift and conformal space projections, not the gravitational equation, rather it is the gravitational equation that resulted from the projections.
(4). Electrodynamics and gravitation are already unified according to the 5D projection theory, neither is there a singularity in the gravitational field solution. The removal of the singularity or the introduction of black holes are not needed, nor do we have black holes in the universe! Totally different to current Cosmology theory .
(5). Both galactic centers and star and planet centers are in 5D, and filled by massless charged spinors closed orbitals, not black holes.
(6). Gluon potentials between quarks are repulsive, but not the result of scattering with Higgs fields in a condensed state, as some conventional theories suggest.
(7). Time reversal symmetry is absent due to the 5D metric not broken by imposing a super symmetry, as specified by the causality requirement, rather it is the inverse.
(8). Stars and planets are formed from projection actions, not successive collisions from asteroids in space as some astrophysicists propose.
(9). The creation of the universe is due to projection actions, not a Big Bang due to massive energy disturbance. As projection has no inverse operation, the universe cannot revert back into pure energy. Furthermore, for the projection actions to create galaxies, it has to start at least a finite time greater than any one galactic 5D core age. This means the age for the universe must exceed the age measured starting with the single Big Bang as now generally believed by astronomers (see recent results, Planck Mission [3]).
(10). All fields are essentially electrodynamic in origin, which always satisfy gauge invariance. We need no other new fields in nature, contrary to what most theories postulate.

## 2. The lepton masses due to boundary conditions created by $P_0$ and the gravitational potential obtained through the projection P and the Riemannian curvature.

In order to fully understand our statements in the introduction regarding the differences between the 5D projection model theory and some current conventional theories, we need an understanding in a geometrical simple way how we map from the 5D homogeneous space-time into the 4D Lorentz space-time as proven by Perelman [4;5], such that we can reformulate General Relativity and develop a galaxy model. To discuss the 5D projection theory beyond producing the hadron representations [1;2], we need to apply the mapping theorem from 5D to 4D as rigorously proven by Perelman. Instead of the rigorous mathematics let us begin by introducing a simple geometrical explanation:
Let us consider a 2D circular coordinate transformation

$$x^2 + y^2 = r_0^2 \qquad (2.1)$$

We can apply a complex transformation with a radius vector r;

$$r = r_0 \, e^{i\Phi} \qquad (2.2)$$

where $r_0$ is the radius amplitude, and $\Phi$ is the angle measured from the x axis.

Now let us define in 3D, that $r_0 = az$, where a is a proportional constant. The 3D surface given by

$$x^2 + y^2 = (az)^2 \qquad (2.3)$$

is a conic surface. It should be noted that the complex transformation for the (x,y) is a conformal transformation. Such surface in 3D breaks the homogeneity of 3D.

Furthermore, the curvatures on this conic surface produce a Riemannian curvature given by:

$$dz/dr = a; \; dy/dx = \tan\Phi, \text{ etc.} \qquad (2.4)$$

Suppose space-time is given by the homogeneous 4d metric: $t^2 = x^2 + y^2 + z^2$, then we make another complex transformation between (z, r) for $r^2 + z^2$ of the 3D manifold. This will give us the surface of a sphere with radius $r_0$. It should be noted that because of homogeneity between all 3 coordinates, despite two rotational angles, this degeneracy of x,y,z reduces the two normals to just one along the radius vector, as any planar cut through the sphere will give always a circle. Furthermore, the electro-magnetic fields given by Maxwell equations indeed is given by this 4D homogeneous metric, and resulted in spherical light propagation from a point source. Thus fixing the center of the sphere by the turning on of the light source, which is the same philosophical conclusion that we stated in describing that the 5D homogeneous space-time metric implies an absolute beginning in creation .

However in 4D space-time with the presence of mass, according to Special Relativity is given by the Lorentz metric, not the homogeneous metric.

Now let us consider the Lorentzian metric, and express it as follows:

$$t^2 - \tau^2 - z^2 = x^2 + y^2 \qquad (2.5)$$

the right hand side can be transformed by the above complex transformation, reducing the

metric to

$$t^2 - \tau^2 - z^2 = r_0^2 \qquad (2.6)$$

As $r_0^2 > 0$, 
$$t^2 > \tau^2 + z^2 \qquad (2.7)$$

If $\tau$ is the projected value of the 4th space dimension from a 5D homogeneous space-time onto the remaining 4D Lorentz variables, then it is obvious that such a projection can be taken only after 5D space-time was created. Consider a conformal projected i$\tau$ vector to be onto a vector r.

Now, treating the fixed vector $\tau$ as a vector independent of z, the Pythagoras sum can again be replaced by a complex transformation to S:

$$\tau^2 + z^2 = S_0^2 \qquad (2.8)$$

where $S = S_0 e^{i\chi}$, $\chi$ is an angle measured from the vector on the straight line joining the +,- $\tau$ values. Here $S_0$ is a variable, while $\tau$ is fixed. Should we consider $\tau$ as the fixed conformal projected value of the 4th space coordinate onto z, then there are two distinct ranges of angles $\chi$: $90^0 < \chi < -90^0$, instead of a circle. It is this $\chi$ angle that breaks the homogeneity between the 3 space coordinates, and give us the problem of proving the Poincaré Conjecture.

Combining two complex transformations, we get

$$t^2 - S_0^2 = r_0^2 \qquad (2.9)$$

This new form of the Lorentz metric, contains two rotational angles $\Phi$ and $\chi$ hidden in the complex vector-variables r and S.

We now do yet another complex transformation to Q between the complex vector S and vector r, we get

$$t^2 = Q_0^2 \qquad (2.10)$$

where $Q = Q_0 e^{i\theta}$, where $\theta$ is again an angle of a circle measured from $r_0$, which lies on the x,y plane and has 2 opposite directions as measured from the center of the doughnut tube, thus producing the two parity states. Hence it is clear that the 3D Lorentzian surface is given by the amplitude $Q_0$, and 3 angle variables $\Phi$, $\chi$ and $\theta$. For each angle we have a plane and therefore a normal. Thus the mapping of the 4D space into 3D space is a 3D surface which has 3 normals (Perelman's theorem [4;5]). If $\tau$ is a constant, then this finite 3D-surface of the 4D volume when t is fixed, is exactly that of a 3D doughnut volume, where the angle $\Phi$ gives the doughnut plane symmetry, while the angle $\theta$ is the doughnut tube angle. The two separated parallel imaginary line-solutions given by angle $\chi$ denotes the two parity states, convex or concave, of the Ricci flow doughnut rings. For the leptonic states we must apply the quantization of $\Delta t \cdot \Delta Q_0 > h$, such that we have the lowest quantum doughnut structure. For the $P_1$ action it is this realization that provides us the foundation on General Relativity and a model for the galactic structure, as matter can only exist in Lorentz space-time, and such a space must be Riemannian. Suppose there is a single $\tau$ fixed value for the universe, then because of the 5D homogeneous space-time, the remaining 3 space variables must all be at least of magnitude $\tau$ before the projection action for the universe can be made. Substitute this into the 5D homogeneous metric, we arrive at the conclusion that the beginning or creation for the universe can only be made at time $t_0 > 2\tau$. In another word the universe must be older than the Big Bang. Furthermore because of this $t = \tau$ value does not imply no projection actions can take place before $t = t_0$, even if there is only one unique $\tau$ value in the galaxy sequential projection action model.

In fact for the elementary particles $\tau^2$ is the Pythagoras addition of the time shift projection, which gives the SU(2) leptons and the conformal space projection that gives the SU(3) quarks. Each part alone is always less than $\tau$. Thus elementary particles must be created way before the possible existence of galaxies with its stars and planets in the universe.

Before going further to tie up elementary particle physics with gravitation let us review some previous work [1;6]. It is informative because of competitive models, such as the Higgs theory. Although the Higgs model can account for much of hadron masses, it does have more difficulties explaining the leptonic masses. First, we know that the quantization of the 5D metric gives rise to the mass-less vector and spinor fields, such that they can be coupled via the unit charge 'e' acting as the coupling constant. Such couplings can be removed by gauge transformations with specific gauge fields depending on the space dimensional geometry of the particle's dynamics restriction, such as in the 2-dimensional Chern-Simons gauge, while a 4D momentum dimension projection action led to the geometric structure of SU(2)xL and SU(3)xL, where L is the Lorentz group. The SU(3) structure gives us one unique bare quark mass, but since there are 4 momentum components in the 5D space-time manifold, hence there should be 3 $P_0$ mass eigenvalue solutions according to Perelman's theorem for the Dirac equation satisfying the 4D Lorentz manifold boundary conditions, leading to the three leptons: the electron, the muon and the tau. One further important consequence of the $P_0$ projection action on the 5D homogeneous manifold is that it breaks time reversal symmetry, therefore decay processes involving the leptons will have to show violation of time invariance, which should be experimentally verifiable [7]. It is our objective to give a simple pictorial explanation here. For the $P_1$ action, which gives us the quarks, and through gauge invariance constraint we obtained the repulsive gluon tensor fields formed from products of the vector potentials generated from intermediate quark currents for the hadrons, such that we can obtain the mass levels in the hadron's SU(3) representations. Since the $P_0$ action gives us SU(2), we found that for the non-vector field coupled spinor solutions, the neutrino, it must be mass-less as its charge is zero, such that the ratio charge over mass can be a constant as we had shown required by the gauge invariance restriction. Hence these neutrino spinor solutions can be obtained when the Lorentz boundary is imposed over the universe. Meaning the neutrino is not coupled to the vector potentials within the Lorentzian universe. The result showed that these solutions possess an oscillation factor because the universe's 3D coordinate manifold can have a geometric form of a doughnut by varying $p_4$, and so does the value of q/m, even as q vanishes, as long as m also vanishes, as shown by the proof to the Poincaré conjecture by Perelman. Therefore there are 3 values for q/m as both q and m go to zero. These 3 values, due to SU(2), must be the same as those ratios given by the masses of the leptons: e/m(e), e/m(mu) and e/m(tau). Thus from eq.(2.8) of ref. [6]

$$A_\mu(R) = \frac{1}{\pi} \frac{\omega_4}{\alpha} R_\mu \qquad (2.11)$$

which means $\omega_4/\alpha$ is a constant. Hence from the 3 alpha values for the leptons we will obtain the 3 $\omega_4$ frequencies, and from which we can verify the validity of our 5D projection model experimentally. To obtain the masses for the charged and massive spinors, the leptons and anti-leptons, we also need to impose the same Lorentz boundary conditions on the Dirac equation in order to obtain these mass eigenvalues, bearing in mind that the micro-5D manifold with the Lorentzian boundary is only over the $P_0$ action impulse time and not that of the universe, because the lepton once formed, its mass is fixed, and the Lorentzian energy-momentum will always be obeyed. For the evaluation of the leptons' rest mass we note that $P_0$ action reduces the 5D homogeneous space-time into the 4D Lorentzian space-time. Since we are interested in the rest leptons masses, it means that we are interested in the perpetual stationary De Broglie waves in the remaining 3D coordinate doughnut form micro-volume they occupy. Due to SU(2), every lepton creation with a net charge e, from pure energy, a massless and charge-less neutrino must also be created to conserve zero spin. As an anti-neutrino does not have charge and cannot be coupled to the vector potential anywhere in the Lorentzian universe it must obey the exact same boundary condition as the neutrino if a solution exist, hence actually it is not a different solution to that of the neutrino. Therefore if it exists and is different to the neutrino, it must be annihilated well within the Lorentzian universe, meaning it is independent of the Lorentzian boundary, in another word this antineutrino does not have boundary conditions. But according to partial differential equation a field with no boundary condition has no solution. It is this asymmetry between neutrino and anti-neutrino that we can have only leptons with negative e charge with its neutrinos in the SU(2) representation. In another word we only have a matter universe with leptons and neutrinos, and no mirror image anti-matter positively charged leptons with anti-neutrinos universe. Meaning anti-leptons can only be created by pair production and not by $P_0$. To analyze the 3 leptonic masses let us try to explain Perelman's theorem in a pictorial way. Consider the 5D metric:

$$t^2 - x_4^2 = x_1^2 + x_2^2 + x_3^2 \qquad (2.12)$$

If we replace $t^2-x_4^2$ by $R^2$ in the left hand side for a homogeneous 3D space manifold, it maps a sphere, and R is the radius. But according to eq. (2.12), the shortest distance between 2 points in 3D space must be less than the distance travelled by light between these two same points implying that light must travel in a curve, when $x_4^2 > 0$, we must have a Riemannian space. Furthermore, according to Perelman's theorem the general 3D connected volume is generally a doughnut structure. The mapping between the sphere and the doughnut is due to the Ricci flow theorem. It should be remembered that the 3 normals to the doughnut structure remain invariant exactly like the radial normal for the surface of the homogeneous sphere. We refer our readers to Perelman's papers for the mathematical proof [4;5]. Thus the field solutions within the doughnut manifold also remain invariant. The 5D metric can equally be expressed in the energy-momentum representation. We get

$$E^2 - p_4^2 = p_1^2 + p_2^2 + p_3^2 \qquad (2.13)$$

Again, if we replace the left hand side by $F^2$, and if F is a pure wave energy, equation (2.13) is the Maxwell 4D metric, which means the wave solution to this metric operator propagates radially at light speed, with an intensity inversely proportional to $R^2$. Hence the normal of light propagation is radial and defines the homogeneity of 3D space. It should be pointed out that the homogeneous 3D sphere does not fix any elementary particle mass based on $P_0$, as $p_4^2$ can be replaced by any m values from 0 to E. Therefore any elementary particle created by P must satisfy Special Relativity. In fact as long as m is not zero in space, a photon with frequency given by $p_4=0$ in space must be shifted in the presence of such a mass located at point x in the 3D space, according to the metric. Again it means the photon path is bended by the presence of m. In other words, we have to unify gravitation with electrodynamics. However if the 3D space is not homogenous, then the most general form of the 3D volume is given by a doughnut by varying the left hand side value in eq. (2.12) as proven by Perelman. In another word, we will have 3 normals rather than just one. It is this multiplicity of the normal to the 3D volume surface and their relative scale that will lead to the 3 choices allowed by the $P_0$ projection. Thus by choosing a set of 3 $p_4$ relative values, we fix the 3 independent normals of the doughnut structure. Corollary, by finding the 3 normals of the doughnut structure, we can determine the ratio between the 3 lepton masses obtained by the $P_0$ projection. This projection then correlates mass to a curved 4D space-time that must be represented by a Riemannian geometry, and thus gives us General Relativity. We leave the rigorous detail mathematics to the future as it distracts from the purpose of this paper. As eq. (2.13) is a quadratic equation, and if we quantized the 3D momentum operator, we could define the 3 normals by 3 independent planes obtained by investigating the e-trino (charged massless spinor [2]) closed loops within the doughnut volume, following the Ricci flow theorem. As a simplistic illustration the eigen-masses can be solved by identifying the 3 normals given by the smallest micro-doughnut volume with 3 standing wave ground state solutions with fixed wave-lengths at the smallest allowed t, given by eq. (2.12), one proportional to the circumference of the doughnut tube $2\pi(r)$, where r<R/2, and R is the radius of the doughnut, as the e-trino that is converted to the leptonic state must be inside the doughnut tube. The other standing waves must then be proportional to the center core circumference, and the circumference through the center of the doughnut tube. As the normal for the standing wave due to the quadratic metric is inversely proportion to the wave-length square, hence it means one of the allowed $P_0$ projected e-trino states' normal is inversely proportional to $(aR)^2$, where a<1, is the scaling needed to adjust for the size of the doughnut center hole dimension. If the higher energy level that represents the tau mass is proportional to the normal to the standing wave in the tube, then the muon mass is lighter by the factor $(aR/r)^2$. Such a ratio remains invariant due to Ricci flow theorem over time. It is easy also to see that these two ground states are not stable if a vector potential field is imposed, since they are fixed only in 2D space inside a 3D domain, and hence can freely move along the remaining axis. However, there is yet another static state, where the wave length is given by a closed loop that wind around inside the doughnut tube completely, pictorially like the smoke particles in a smoke ring. Hence its wave length is proportional to $2\pi(r+R/2)$. This wave is fixed in 3D, and therefore remains stable when a vector potential is imposed. As the imposed vector potential energy increase, it can be transferred into increasing the amplitude of the wave, which is mainly along the ample doughnut plane, meaning creating more states. Hence if this represents the electron mass level, the electron is stable, and then the mass ratio between the electron and the muon is given by $(r/r+R/2)^2$. It is easy to see that r and R are of the same order of magnitude. In fact the doughnut domain confined waves can be treated as the interpretation of a lepton's de Broglie wave. While for the hadrons, it is the gauge loop confinement on the quarks constituents that produces the hadron's de Broglie wave. Let $r=nR$, $n<1/2$, measure the ellipticity of the doughnut tube. Note that the trapped muon could move marginally horizontally within the tube as the elliptic height narrows away from the tube center but not up or down at all, which means the e-trino wave is pinned at the top and bottom of the tube, but it can oscillate horizontally as well as go around the tube. In fact this e-trino wave form for the electron state is a simple single twisted topological oscillating wave. With this model then the electron mass/muon mass is given by $1/(1+1/2n)^2$, while the muon mass/tau mass = $(a/n)^2$, such that $a/n<1$. In fact we can get good fits to the lepton masses, even though the standing wave model is crude. The values for $a=1/110$ and $n=1/27$ give us reasonable good fits to the experimental lepton masses m(e)=0.5MeV, m(mu)=105.2MeV, m(tau)=1745MeV, although errors associated with ignoring the core exclusion as well as the elliptical shape of the spinor waves, with nodes on the top and bottom of the doughnut tube, and the maximum amplitude in the doughnut plane may produce errors of a few percent, but because of the small core and the uncertainty principle, the leptonic masses obtained from ignoring the core size might be justified and accurate. The core can never be of zero size, as the tau lepton obtained from $P_0$ must have a finite horizontal oscillation amplitude. None the less this crude model produces a geometric shape of a ring like flattened doughnut manifold, with a sizable but relatively small center core, and this geometric picture governing the $P_0$ creation of leptonic masses in fact resembles the mass distribution picture of a galaxy. Again presenting an illustration of fractal dynamics, from the micro to the macro systems in nature. If we ignore the core aR, then the tube ellipticity 'E'=2n=2/27, a gluon strength factor that is found in the baryons. We like to point out that this numerical resemblance does not mean that leptonic physics also obeys the standard model of the hadrons. It is interesting to observe that from the standing wave form of both the electron and muon states, their motions must be spread inside the 3D tube volume of the flattened ring, while the heavy tau is confined to a 2D loop around the central core. It should be remembered that the lowest level wave length, say for the muon is around the tube 1/2 way inside the circumference, thus it is given by $2\pi ER<R$. The importance is that their motions are in 2D perpendicular to the doughnut plane. Hence an alternating magnetic field perpendicular to the doughnut plane will induce a transition of the tau or muon lepton to decay into lower mass leptonic and anti-leptonic states, with emission of photons and its corresponding leptonic neutrinos. Since only the electron state is stable, we would expect an extremely relativistic electron state at least of an energy much larger than the muon rest mass, as it propagates through the universe it must be also expressed in terms of superpositions of the eigenfunctions which consist of all three leptonic functions that satisfy the universe's boundary conditions. In another word, it should produce a similar oscillation evolution feature that is detectable like the neutrino? Or maybe we should interpret this as the leptonic law of isospin conservation? One last point we need to consider with the Po projection action. At all time in the 5D homogeneous manifold, we need to have no net charge and current. This requirement is important to arrive at the 3 negative charged leptonic eigenstates inside the 3D doughnut space structure. First around the doughnut core, in order to have zero current, the tau leptonic standing wave must be given by a Cooper pair. If we assign the charge to the e-trinos Cooper pair that forms the tau Cooper pair as negative, then it follows, the $P_0$ action produces 2 negative taus in a Bosonic form of a Cooper pair. Thus in order that the net total charge is zero, we must leave behind in the 5D core 2 anti-e-trinos. Hence breaking charge symmetry in leptonic creation. To understand this, we review the $P_0$ projection on the 5D homogeneous space-time which gives us the proper time $\tau_0$. For the remaining 3D space to have zero volume the time t has two roots, $t=\tau_0$, $t=-\tau_0$. Obviously the root $t=-\tau_0$ is not allowed due to the condition that time only increases as given by the 5D metric. Hence the fact that leptons are only negative is a consequence of no time reversal symmetry in the CPT invariant Lorentzian universe. Hence in order to satisfy charge neutrality, massless anti-e-trinos must be simultaneously created by $P_0$ in the 5D domain. These anti-e-trinos left in the galactic core thus produce the giant gamma ray bubbles above and below the Milky Way galactic plane [8].

As leptons and the quarks, namely the hadrons, are interactive through their common coupling to the electromagnetic vector potentials, therefore in matter, two types of structures are now possible, according to the 5D projection model. The most common is the spatially separated form, like the atoms, where $P_0$ and $P_1$ produce masses that occupy spatially separated domains. As mentioned in [1], namely that a lepton-lepton Cooper pair obtained from lepton-lepton head on collision will affect the mass renormalization similar to that of the p-p collsion, except that this second order perturbative correction for the leptons in 3D space is much smaller. This is because for the leptons gauge constraint is automatically satisfied, therefore there is no leptonic gluon potential, hence the potential field is only the single vector potential, making the flux quantized loop radius of macroscopic dimension possible, similar to the case in superconductors, even if there is a net binding potential between the leptons, if there is no extra attractive potential this corrective term is repulsive, and the h/2e quantum flux does not exist, unlike the microscopic dimension which must be the case in the baryons, because of the confinement needed to keep the 6 quark constituents within the baryon size. However, with further imposing space dimensional restriction, such as the electron motion confined to 2D, for example the Chern-Simons gauge would apply which leads to a collapsed bound state of the hydrogen electron, (see [9]) forcing the electron's ground state into the inside of the proton nucleus. Furthermore, this hydrogen state has an anyon statistics. It is this amazing result that led to the possible presence of leptons and their accompanying neutrinos inside the nucleus, hence certain nuclei will give beta and neutrino radiations, as well as producing the Z, W leptonic bound states in nucleon scattering experiments. With this quantum picture, the heavier muon and tau leptons would necessarily decay into the stable electron by photon emission and accompanied by an oscillating phase shift on the accompanying neutrino (see ref. [6]). As discussed in [1;2], the general dimensional reduction projection allows for the additive $P_0$ and $P_1$ actions, where $P_0$ action produces parity violating representations (see [2]), because the 5D homogeneous space-time is not time reversal symmetric. In another word the 4D Lorentzian universe always expands with increasing time, such that the contracting universe solution is not an acceptable choice. However the $P_1$ conformal space-space projection is time invariant because time is not involved. Hence the addition of both $P_0$ and $P_1$ actions can be employed as a model for the creation of the non-reversible Lorentzian universe, like a Big Bang, which of course must include the creation of planets and galaxies.

Kaluza (1921) first discovered that the Newtonian Gravitation cannot be derived without extending the 4D Lorentz Space-Time to at least one higher dimension if we want to combine gravitation to electro-magnetism, as we have discussed in [1;2]. However he ran into the problem of designing a mathematical consistent way to get back to the 4D Lorentz space-time. By assigning the gravitational potentials $g^{\mu\nu}$ also as the metric line element $ds^2 = g^{\mu\nu} dx_\mu dx_\nu$, Einstein (1938) obtained the field equations in General Relativity. As we discussed earlier in the chapter, by applying a projection P on the 5D metric, we produce both a mass and a proper time, which gives us a Riemannian curvature and thus the metric line element. However, P can also produce the Maxwellian metric, when mass and proper time are both zero. Hence a unified gravitation and electrodynamics theory can be constructed by a linear superposition of the two projection results on the line element.

On carrying out the line element integration over the 5th dimension [10]

$$\delta S = \delta \int d^4 x R_4 F g + \delta \int d^4 x \{-\tfrac{1}{4} F_{\mu\nu} F^{\mu\nu}\} F g \qquad (2.14)$$

where $R_4$ is the invariant derived from the 4D curvature tensors, and g = det. $g^{\mu\nu}$, which yields half of Maxwell's equations along with the Einstein gravitational field equations. We refer to reference [10] for details on the derivation and notations of eq. (2.14). By assumption the 5th dimension has the metric for a covariant line element given by $\gamma_{\mu\nu}$ for $\mu = 4$, or $\nu = 4$. When Kaluza's work was generalized one could see that these line elements $\gamma^{\mu\nu}$ were related to the electro-magnetic vector potentials by

$$g^{\mu\nu} = \gamma^{\mu\nu} - A^\mu A^\nu \qquad (2.15)$$

Einstein made the electro-magnetic potentials depend on the 4th space coordinate periodic, thus 'closing' this dimension, so that it curled around itself, and not stretched out in space. This is of course different to our 5D homogeneous space-time assumption. In Einstein's case, the line element takes the form

$$ds^2 = g^{\mu\nu} dx_\mu dx_\nu - (dx_4 - A_\mu dx^\mu)^2 \qquad (2.16)$$

The $A_\mu$ can be interpreted as coordinate transformations in the 5th dimension, as could a fifth potential, $\Phi$ be inserted into the expression of eq. (2.16). (see ref. [11] for more discussions). Thus the electromagnetic potentials have a dual reality, serving also as a coordinate transformation in the 5th dimension. The coordinate transformations now take the form of a gauge transformation,

$$g'_{\mu 4} = g_{\mu 4} + g_{44} \partial_\mu g \qquad (2.17)$$

which implies $A'_\mu = A_\mu - \partial_\mu f$

In other words, gauge freedom is a geometric freedom in the 5th dimension.

For the homogenous 5D space-time and the projection action Einstein's assumption cannot be made in general, although we must also find a way to make our projection model produce a Riemannian Geometry in 4D Lorentz space-time, such that via some closed path integral, like a gauge transformation, the gravitational potential is a result. We should point out that the solution to the Maxwell metric operator for a scalar field will produce a Coulomb form potential, except that we cannot obtain the Newtonian constant from a mass point source.

Before establishing our model for gravity, we should understand what the space-time projections mean. We start with the time shift projection $P_0$. Explaining in the context of the first interpretation for the 5D homogeneous space-time metric, $P_0$ implies we convert this 5D homogeneous space-time into the Lorentz space-time at the instant $t = \tau_0$, after the creation of the 5D space-time manifold. The conformal projection $P_1$ has a totally different meaning, as it changes any line vector into a complex vector. This corresponds to a circular transformation. For example we transform a vector in the x, y plane into the circular r, θ representation. By allowing a continuous change of the angle θ, it will lead to a circle. In the projections dealing with charges, due to gauge invariance constraint the phase factor is discrete, resulting in a SU(3) symmetry, rather than a polar transformation. Since for the momentum projections for the creation of mass we have no gauge constraint, $P_1$ can generate a continuous polar coordinate transformation, starting with an initial radius $c\tau_0$, as long as the Lorentz metric proper time τ, satisfies

$\tau^2 = \sum_{\mu=0}^{3} \tau_\mu^2$. This restriction leads to two possible geometrical mass distribution forms in

the 3D space, namely a spherical shell mass, or a doughnut mass distribution. It is interesting to point out that for the doughnut mass distribution in 3D the doughnut core is in 4D at time t after the beginning. The 3D manifold can be mapped into a solid 3D sphere with a 4[th] dimension line passing through. Since the 4[th] dimension is finite as defined by t, it must be closed, thus this geometry can be mapped into a 4D symmetric shape. In another word, satisfying Poincaré Conjecture. This mapping implies that the doughnut model can be mapped into Einstein's spherical model. These geometrical forms alone are not sufficient for obtaining the Einstein eq. (2.14). To achieve that we need the Riemannian curvature

$$g^{\mu\nu} = \frac{\partial x^{\mu}}{\partial x^{\nu}} \qquad (2.18)$$

where μ and ν run from 0 to 3.

Take the cylindrical form as depicted by the doughnut mass structure. The geometry gives us $\partial x/\partial y$, around the 3D core which can be converted into circular polar coordinates. We still need $\partial z/\partial r, \partial z/\partial t$, and $\partial r/\partial t$. These Riemannian curvatures are related by the ratio between the rates of radial and vertical mass creations enacted by the projections. For the spherical structure the rate of mass creation is not needed, because of symmetry. With these curvatures defined we obtain the Einstein equation without Kaluza's assumption of tying the curvature to the electro-magnetic vector potentials as given by eq. (2.15). Similar constructions can be made based on the spherical mass shell geometry. Obviously the projection approach eliminates the necessity of requiring the closed loop constraint on the 4th space dimension as Einstein assumed in General Relativity in all mass form geometries. Although the symmetric form of the spherical shell will necessarily produce a curvature similar to the Einstein curvature when it is required that outside the core we have only the 4D Lorentz space-time. Hence this is equivalent to requiring the 4th coordinate to be closed as assumed by Einstein, and therefore it corresponds to the compactification of the 5th dimension. Compactification of the 5th dimension is the source of singularity in General Relativity and an energy source for the projection spherical mass shell model. We borrow from the ideas of Viswanathan [12]. As the 5th dimension is closed, any mass-less fields that can exist must be periodic within that dimension. Therefore as this path length L reduces to zero, the field's lowest frequency 2π c/L becomes infinite, leading to the gravitation singularity in General Relativity. On the other hand, the projection model removes the zero length limit, such that for this geometry there would exist a mass empty 4D core. Such an empty core will remove the existence of a Schwarzschild singularity for any mass value for the sphere [13]. Hence it is corresponding to our proving of the existence of a worm-hole inside. Furthermore, the solutions for the e-trino fields must be expanded in the 4D periodic standing modes. As they carry a charge, they produce electro-magnetic energy. Since the spherical geometry model being totally similar to Einstein's, its Riemannian curvatures can be added to the electro-magnetic potentials producing

$$g^{\mu\nu} = \gamma^{\mu\nu} - A^{\mu}A^{\nu} \qquad (2.19)$$

In other words, we have proved that gauge freedom becomes the geometric freedom in the 5th dimension, which definitely also holds for the projection model.

The presence of mass-less empty cores in these mass distribution geometries raises the question of their stability, as masses attract, which would produce an inward pressure on the structures, unless compensated by an equal and opposite outward pressure. The presence of such outward pressures in the 3D mass-less core is a natural consequence of the 5D fields, which we have discussed in [2;6]. To illustrate this point, let us again consider the doughnut mass distribution structure. Energy inside the 5D cylindrical core can create pairs of opposite momentum and charged e-trino and anti-e-trino. Such an opposite momentum pair is confined to travel along the z direction of the cylinder and the 5th dimension. Due to their opposite sign charges, they will produce a net current along z. A net z current with an average frequency due to the e-trino pair energies produces a circumferential magnetic radiation field as well as the z direction electrical radiation field. Therefore it will exert an outward electro-magnetic pressure on the surrounding masses of the core. We will discuss this feature further when we apply it as a model for the creation of a galaxy in Section 3. It is not difficult to come to a similar outward electro-magnetic radiation pressure for the spherical mass shell structure. The e-trinos travel at light speed in all directions of the 4 coordinates, within the 5D homogeneous space-time domain. Thus the mass shell enclosing the 5D void serves as a closed boundary to the e-trinos' motion. In fact, this 5D void could be seen as actually freezing the time to tau, or in quantum language making the e-trinos' states into time independent standing waves within the 3D void of the core. These standing e-trino waves and closed loops would produce permanent magnetic fields. Thus if a spinning motion of the mass shell is present, and if applied as a model for the formation of stars and planets, (see [1]), then this magnetic field will be in the form of a dipole field, similar to objects in the solar system.

As for the doughnut geometrical structure, because e-trinos are mass-less, they have energies and momenta similar to photons. They will contribute to the sources in the 4D Maxwell equations. Since the hole has open ends, the e-trinos are not trapped. Not only can they escape along the 5th dimension, they also provide extra source terms. Other than an addition to the scalar charge density for the Coulomb potential term, they produce an additional current term of strength

$$j_4 = ce\bar{\Psi}\Psi \qquad (2.20)$$

along the z axis of the doughnut structure. $\Psi$ is the e-trino spinor.

Since the e-trino velocity is 'c', this additional term resembles that of a magnetic monopole as suggested originally by Maxwell [14]. Solving the electro-magnetic fields with such an additional source makes the solutions for the E and H fields symmetric. We refer our readers to Maxwell's original paper [14]. Despite overwhelming experimental results to the contrary at that time, Maxwell still insisted on the symmetric electromagnetic theory, meaning that his 5-component vector potential theory was really amazing.

## 3. The 5D to 4D projection model applied as a Lepton and Galaxy Creation model.

In Section 2 we showed that the mixed $P_0$ and $P_1$ projections that resulted in a spherical 3D mass shell with a mass void core actually produces the confinement of the 5th dimension within the core, as the space-time outside must be just the Lorentz space-time. This condition thus satisfies Einstein's assumption by making the $5^{th}$ dimension closed, and will produce Einstein's Riemannian curvatures, hence leading necessarily to the Einstein gravitation equation. Except because of the existence of a finite mass void core we have a build-in worm hole of radius $\tau_o$ and a mass shell of thickness $\tau - \tau_o$, which will prevent the existence of a Schwarzschild singularity inside the spherical mass structure. It was also possible to include the electro-magnetic field tensor to the gravitational tensor in the spherical structure from which we get the connection of geometry freedom and gauge invariance. Furthermore, because the 5D is homogeneous, if the $4^{th}$ space dimension is closed onto itself so must the remaining 3 space dimensions be. Thus the e-trino spinor fields inside the void must also only propagate in closed loops. This is of importance for the production of outward electromagnetic pressure acting on the enclosing mass shell, as we have discussed earlier in [1;2]. It is easy to show in the spherical representation that we have the following Riemannian curvatures: $dr/dt$, $rd\theta/dt$, $rd\Phi/dt$, and $d\theta/d\Phi$, only. Obviously due to symmetry these curvatures are related to the rate of mass creation by projection. For example $dr/dt = (dm/dt)/(dm/dr)$, while $d\theta/d\Phi = 1$. Thus it is straight forward to show that $\sqrt{-g}$ gives the Newtonian constant.

There is no difference here between the projection model and general relativity except that the projection model presents an action for generating the Riemannian space. However, the projection model can also give us a cylindrical mass-less void structure. In this case, the $5^{th}$ dimension is not compacted. In fact it is allowed to extend to the boundary of the universe as given by the homogeneous 5D space-time metric. We have mentioned in [1], that the Riemannian curvature from such a doughnut mass creation model depends on the ratio of the rate of mass increase along the cylindrical axis to that along the doughnut core radius. This ratio is not arbitrary as the gravity factor 'g' = det $g^{\mu\nu}$, must be independent of the structure so that it is the same as that obtained from the spherical model. We will try to describe the process in a non-rigorous way, yet still provide us a physical picture, as we want to employ it as a Galaxy model.

The Riemannian curvature is given by

$$g^{\mu\nu} = \frac{\partial x^{\mu}}{\partial x^{\nu}} \tag{3.1}$$

with $g^{\mu\nu} = -g^{\nu\mu}$, where $\mu$ and $\nu$ runs from 0 to 3.

Let us consider a cylindrical structure for the 3D coordinate space. The structure in the x,y plane obviously is circular. There remains the curvature between z and the circular radius r. We note that we can define

$$g_\varphi = \left(\frac{\partial z}{\partial r}\right)_\varphi = -g_{-\varphi} \qquad (3.2)$$

where the angle φ is clockwise, hence counterclockwise means the variation of -z with respect to r. Under projection, a constant amount of mass is created as a function of time t. We therefore expect a radial rate $\Gamma_r$ and a z direction rate $\Gamma_z$. As mass is created its distribution size will expand proportionately. Therefore

$$g_\varphi = \left(\frac{\partial z}{\partial r}\right)_\varphi = \frac{\Gamma_z}{\Gamma_r} = \frac{1}{G} \qquad (3.3)$$

According to the projection action, we start at $t = t_g = \tau_0$. Thus the circular base radius starts at $c\tau_0$. The Lorentz invariance requires $\tau^2 = \tau_0^2 + \vec{\tau}^2$. Therefore it implies that we have a continuous variation of the cylindrical core radius from $c\tau_0$ all the way to $c\tau$. This means that the core consists of two conic structures stacked opposite to each other. [see Fig. 1] The slopes of these cones are given by $g_\varphi$.
This curvature is the result of the Lorentz invariance, similar to the Einstein case.

Now for this structure, we have $\partial z/\partial r, \partial r/\partial t, r\,\partial\varphi/\partial t, \text{ and } \partial z/\partial t$. While $\partial\varphi/\partial z = 0, \partial r/\partial\varphi = 0$. Constructing the det $g^{\mu\nu}$ gives the gravity constant, and the gravitational potential equation which is equal to the simple description below.

Suppose we integrate around a circular loop with radius 'r' from the center of the core, larger than $c\tau_0$, and enclose a total mass 'm'. Then from equation (3.3), in place of Kaluza's gravitational equation, we will obtain a gravitational potential Φ generated by the mass m, as if it is a point mass located at the center of the core, satisfying:

$$\oint d\varphi\, r\, \Phi\, g_\varphi = \pm 2\pi m \qquad (3.4)$$

This non-rigorous method cannot fix the sign of the potential, mainly because the doughnut structure is not reducible into a homogeneous sphere with no preferred axis. In another word, the Poincaré conjecture problem. With proper and careful mathematical mapping consideration this problem can be removed. Similarly if the loop is inside the core, then m = 0, and the gravitational potential vanishes.
Note that for this structure the 5th dimension is not integrated, and is extended along z inside the core to the boundary of the universe.

As mentioned in section 2, the creation of e-trinos and anti-e-trinos from energy inside the cylindrical mass void core, can produce a net current along z, which in turn will produce an outward electro-magnetic pressure on the masses outside, which has no electro-magnetic pressure. In hydrodynamics, a pressure gradient configuration will produce an anti-cyclonic flow of a fluid. In this case the fluid is composed of mass particles. The fact that a pressure gradient always produces circular motion is due to

Newton's 3rd law. Circling motion of a mass possesses angular momentum and centrifugal force so as to counter act the radial pressure gradient. For example, water inside a sink will form a circular flow inward into the drain, due to a negative pressure gradient. In this doughnut mass distribution case the pressure gradient is positive. Therefore the circular flow is outward. This picture certainly resembles star motions within a galaxy.

Suppose we do accept this structure as the galaxy model. Let us consider the motion of one star, mass M, outside the cylindrical core. Suppose we can describe its motion classically. Its equation of motion along the radius r from the core is given by

$$M \frac{d^2 r}{dt^2} - \frac{L^2}{Mr^3} = F \tag{3.5}$$

where L is the angular momentum, F is the radial outward force produced by the electro-magnetic radiation pressure from the core e-trinos. The factor F falls off with increasing r. It is qualitatively obvious, that the star's outward acceleration is very high when it is close to the core. As it spirals outward, and if angular momentum is assumed conserved, then both the centrifugal force and the force from the electro-magnetic radiation will fall off, leading to a steady outward velocity. Because of the presence of F, angular momentum is not a conserved quantity. It is more accurate to assume a relatively constant angular velocity $\omega$, at least during the early stage of the star. Under this assumption, equation (3.5) becomes

$$M \frac{d^2 r}{dt^2} - \omega^2 M r = F \tag{3.6}$$

The solution of r to equation (3.6) with F assumed constant, is

$$r = r_0 + (\tfrac{F}{M\omega^2} - r_0)(1 - e^{-\omega t}) \tag{3.7}$$

The initial value $r_0$ is roughly equal to $c\tau_0$, and t is the time after its creation.
It is the exponential t dependence of r and the radial velocity that is interesting, and could be observable [15].

$$\frac{dr}{dt} = (\tfrac{F}{M\omega} - r_0 \omega) e^{-\omega t} \tag{3.8}$$

The factor F/(M$\omega$) in the solution for dr/dt in equation (3.8) lead to the existence of the so called run-away stars during the early stages after galaxy formation. However they should not exist among those near the galactic edge. By the notion run-away [16], we mean that its speed is increasing much faster as compared to other stars that are moving close to their steady orbital state.

By adopting the doughnut model for the galaxy, we also have defined the mass void galactic core. This core, although it contains no masses, it is full of energy consisting of electromagnetic energy and e-trino energy. Although it is frequently called a 'Dark Matter' core, it should be very bright, due to the electro-magnetic radiation coming out. In Section 2, we showed that energy in the cylindrical core can produce e-trino and anti-e-trino pairs. These mass-less charges travel along z, with opposite momenta hv/c, thus giving rise to a net z direction current, given by eq. (2.20). Such a current will generate both an electric field along z, which must have both a steady field as well as an oscillating z component, with frequencies ν derived from the e-trinos, as well as a steady circular H field, together with an oscillating part. As the galactic core is open ended, the 5D void will continue to increase in 4D space volume according to the 5D metric, thus the e-trinos and anti-e-trinos inside would radiate. It is the oscillating E, H, components that will produce the intense light and gamma ray emissions in the z direction, which we can see as intense light and gamma ray emissions in the z direction [8], the giant gamma ray bubbles from the galactic center of the Milky Way.

The constant E, H amplitudes are the principle contributors to the outward electro-magnetic pressure on the stars around the core, creating the reverse cyclone motions of the stars, as we described earlier. As long as energy is available inside the core to produce the e-trinos and anti-e-trinos, the stability of the galaxy will be sustained.
The geometrical space volume occupied by the mass-less core can be roughly calculated.

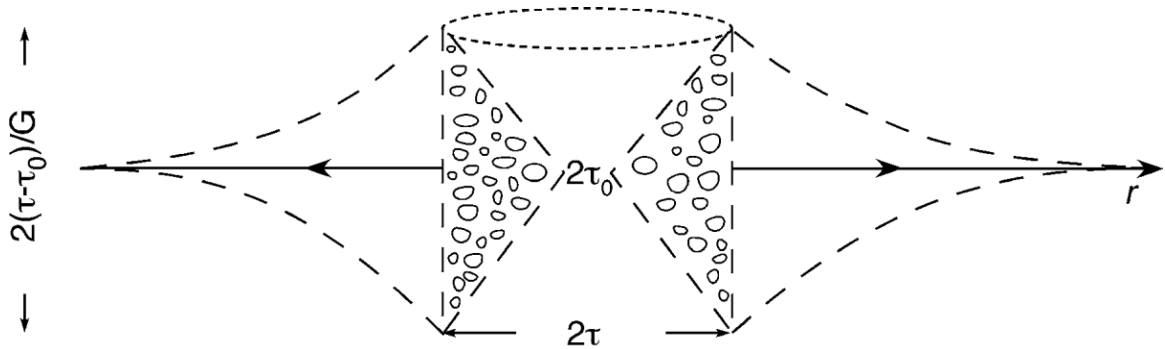

Figure 1. Doughnut structure galaxy model in cross section depicting a time-extended projection action. ($\tau_0$ varies from an initial minimum value to $\tau$ as a function of h, the thickness of the galaxy).

Let us consider the structure as shown above in Figure 1. The cross section of the core varies from $2\tau_0$ to $2\tau$, over the time extended projection action. Starting initially at t = $t_g$ = $\tau_0$, after the creation of the 5D space-time, and ending at t = $\tau$. Ignoring the change in the created mass distribution during the entire action time $\tau$, and assuming a mass creation rate ratio between the rate along the z axis of the cylinder $\Gamma_z$ and that of the radius r in the cross section $\Gamma_r$, such that $\Gamma_z/\Gamma_r = 1/G$, we obtain the cylindrical

height
$$h = 2(\tau - \tau_0)/G. \qquad (3.9)$$

Thus the 3D space volume occupied by the dark matter core V(DM) is given by

$$V(DM) = 2\pi\tau_0^2(\tau - \tau_0)/G + \pi(\tau^2 - \tau_0^2)(\tau - \tau_0)/G \qquad (3.10)$$

while the phase space volume occupied by the created masses V(M) is equal to

$$V(M) = \pi(\tau^2 - \tau_0^2)(\tau - \tau_0)/G \qquad (3.11)$$

Because of Liouville theorem, V(M) is invariant even if the masses are redistributed due to motion. Hence the ratio between V(DM) and V(M) is also invariant and is always greater than 1.

$$\frac{V(DM)}{V(M)} = \frac{1 + (\tau_0/\tau)^2}{1 - (\tau_0/\tau)^2} > 1 \qquad (3.12)$$

Although this analysis is based only on one galaxy, it is not surprising that the data obtained by the Hubble Deep Space Probe showed a ratio of V(DM)/V(M) as much as 1.3 in the universe. These observed astronomical data appear to support our projection doughnut structure galaxy model (Figure 1). As the 4D space expands with time the universe will have larger and larger so-called dark matter space. Such expanded 4D space might not contain e-trinos and would indeed appear dark without electromagnetic fields generated.

In summary, (I) *Gamma-Ray Bubbles*: If there is coupling between two vector potential fields (5D vector potential and 5D e-trino) which are solutions of the 5D homogeneous operator, then their product is also a solution and their multiplication constant between them is their coupling constant. The coupling can be transformed away by gauge transformation. This gauge transformation produces the 4Dx1D Hilbert space representation for the 5D vector fields (see [2]). Hence, in 5D we can have charged source terms for the 5D vector potentials (see [2]), similar to the Maxwell potentials in the 4D Lorentz space-time. This means it shows explicitly the creation of E, H fields (electro-magnetic fields) by the e-trino, as well as the separate monopole potential field (see Section 2), which must be compacted in the 4D Maxwellian space-time. It is this requirement that produces the gamma-ray spherical bubbles in the Milky Way galaxy.
(II) *Galaxy Creation*: By analyzing the initial creation of the 5D space-time manifold, and since it must obey uncertainty principle, a near infinite amount of 5D energy and momentum fields must be created during the initial dt. This initial energy and momenta, when projected into 4D Lorentz space-time later (any time later, at any coordinate space point), can start the formation of a galaxy, by creating a Riemannian Space. Such a galaxy creation is like boiling water in a pot. The size of the galactic core can have different sizes, except the Riemannian curvature determined by the rates of creation must obey the gravitational constant, due to covariance of the resulting Lorentz space-time, as given by General Relativity. The galactic creations do not change the total energy in the 5D universe, as it is simply a projection process in converting one momentum component into mass. In other words, this projection action, hence creation of a galaxy, can happen at any region of the 4D space, at any time t larger than zero. However, it cannot occur in any region which is already 3D space, meaning it is already Lorentzian. Thus the cores of galaxies cannot overlap. Furthermore, because of the restraint on the Riemannian curvature, the core of each galaxy must be invariant. That means the 3D volume size of the 5D core, including the gamma ray bubbles, is a constant, although its shape can change according to Liouville theorem and Poincaré conjecture. No matter how many galaxies are created the total energy and momenta in the universe is invariant. Therefore, ensemble theory and laws of thermodynamics, which are the consequence of ensemble averaging, is valid for any subsystem in the universe, since the universe acts like the thermal bath and is a closed system. The star distribution of each galaxy is also a result of statistics, no different than any gas volume. As to the elements in the galaxy, that is dependent on the stars' planetary structures, which is also a result of statistics and not a fundamental rule. The 4D space regions, as the homogeneous 5D space-time expands, are the residual of coordinate space not already converted to 3D due to projection actions. This portion is always much larger because it has an extra coordinate dimension. The creation of galaxies makes the distribution of 5D fields non-uniform in the universe. Volume not occupied by mass distribution, or Lorentz space, will be dark matter space. The 4D space expansion always increases faster than the reduction due to projection action, and the universe energy and momenta densities per unit 4D volume also always decreases with time. Although according to our projection model galactic cores cannot overlap, galaxies can still collide when they get within gravitational range of each other, which is part of astronomical observations.

(III) *Sequential Galaxy Creation*: We have shown the creation of a galaxy based on time ordered projection actions. In effect or by definition we have chosen the specific beginning at $t = t_{g_1} = 0$ for the start, as if at the beginning of the creation of the 5D homogeneous space-time, and $t = t_{g_1} = \tau_0$, as the narrowest radius of the cylindrical void core, and $t = t_{g_1} = 0$ must only relate to the beginning of the nucleation time of a galaxy at the perimeter of the finite 5D space-time manifold which is defined by the absolute time t. At any time $t = T$, in fact, more than one galaxy can be created at T, displaced from each other in 4-dimension space. It should be noted that according to our galaxy creation model, all galaxies must start from absolutely nothing, no time and no 4D space. Therefore galaxies can be created in sequence after the creation of the 5D space-time, or simultaneously. If all galaxies in the Universe were created simultaneously, then our 5D projection model can be made consistent with the Big Bang model. As to the sequential galaxies creation, let us consider a second galaxy ($g_2$) created with its center starting at $t = t_{g_2} = T$ outside the 5D space-time domain according to the 5D space-time as measured from the first galaxy ($g_1$) which is placed at $t = t_{g_1} = 0$ within the universe, and with its center translated from the absolute t frame. That means that it could be just on the outer edge of the first galaxy. Let $\vec{\tilde{x}}'$, t' be the 5D space-time as measured from the beginning of the newly created galaxy. In this case the 5D space time for the total universe consisting of the two galaxies would be given by $\vec{\tilde{X}}^2 =(t'+T)^2 = t'^2+ T^2 +2t'T= \vec{\tilde{x}}^2 + \vec{\tilde{x}}'^2 + 2|\vec{\tilde{x}}||\vec{\tilde{x}}'|$, where $t'^2= \vec{\tilde{x}}'^2$. $\vec{\tilde{X}}$, $\vec{\tilde{x}}$ and $\vec{\tilde{x}}'$ are 4D space vectors, and because of that the combined two galaxies must remain in a single 5D space-time manifold, $\vec{\tilde{X}}$ is necessarily equal to $\vec{\tilde{x}}+\vec{\tilde{x}}'$, as $\vec{\tilde{x}}$ and $\vec{\tilde{x}}'$ must be parallel, which means this newly created galaxy lies along the path of the growing universe of the original galaxy. Hence this newly created galaxy is observable from the first galaxy. As this galaxy gets created by the projection of the 4th component momenta of the 5D fields into masses around its cylindrical void core which expands from the narrowest radius $\tau'_0$ to the largest radius $\tau'$, it must push away from the outer boundary of the 4D coordinate space domain established by the original galaxy as its core is completed. Newtonian conservation laws requires action and reaction are equal and opposite, therefore setting in a relative drift apart velocity between the two galaxies. As the stars of the newly created galaxy gradually spiral outward from its core expanding the galactic domain, it will further increase the speed of the relative drift apart. Obviously, we can apply the exact same projection action sequence and create the extended Riemannian space so as to cover the masses created within the new galaxy. In the mathematically non rigorous static model we discussed earlier we obtained the Newtonian gravity potential by considering a static initial galactic mass distribution model as depicted in Figure 1, but because as soon as masses are created right after time $\tau_0$, these masses will start to spiral outward due to the tremendously energetic 5D fields caused by the quantum uncertainty principle upon the galaxy nucleation out of nothing we have discussed in ref [1;2], which therefore will produce a pressure already within the planar galactic void core radius $\tau_0$, even before all the mass creation was completed, hence for mathematical rigor the resulting Riemannian Geometry must be formulated into a covariant 4D Riemannian space-time so that mass motions which obey Special Relativity are always satisfied. Therefore the galactic masses can only be taken properly into account by generalizing the Einstein General Relativity Riemannian formulation. One such method is by using Perelman's inverse mapping and changing Einstein's 5th dimensional compactification in the mass volume to the Ricci flow through the doughnut core of the galaxy extracted from his proof of the Poincaré Conjecture. The detail for this method is very mathematical and beyond the level intended for this paper. It should be pointed out that the origin of the 4D coordinate manifold need not be the center of the first galaxy created. In fact a finite 4D space manifold might be formed before the projection actions to the 4D Lorentz space-time. As to the question on the changing number of galaxies in the Universe, there is no requirement that new galactic creations need to continuously occur. Thus in the universe the number of galaxies can be constant? If true then it would also be impossible to distinguish between the 5D projection model and the Big Bang Model. The question is can we still be able to distinguish between our 5D projection theory and the Big Bang? Because each galaxy contains a core of radius $\tau_0$, and it must form on the boundary of the 4D finite space at the absolute time T, which means T must be greater than $\tau_0$, if even just one galaxy was created first at T. However if N galaxies were simultaneously created, then $2\pi T$ must be greater than $2N\tau_0$, where $\tau_0$ represents the average core size of all N galaxies. Thus the age of the universe is much older than the first galaxy creation. This conclusion is not generally recognized by astronomers. Finally, the galactic cores contain e-trinos, and therefore radiate, only the gap spaces between the N galaxies first created can be truly dark space. Thus such dark spaces in the universe might be much less than we might expect. The distance measured from one galactic center to the next although is measured in a curved 4D Riemannian space-time, but by taking into consideration the distance due to their relative drift, we can still get a rough estimation of the age difference between the galaxies, however we cannot determine which one is older, unless we can find the location of the absolute center of the 4D space in the ever expanding homogeneous 5D space-time of the Universe. Except of course as galaxy creation continues the newer galaxies at the perimeter of the 4D space would be still be in the forming stage and therefore will contain new stars that are in the process of being created.